\documentclass[prd,twocolumn,showpacs,floatfix,amsmath,nofootinbib,amssymb,floatfix]{revtex4}
\usepackage{graphicx,color,dcolumn,booktabs,bm}
\usepackage{longtable,lscape}
\usepackage{txfonts}
\usepackage{overpic}
\usepackage{amssymb}
\usepackage{indentfirst}
\usepackage{feynmf}   
\usepackage{slashed}  
\usepackage{cases}
\usepackage{color}
\usepackage{multirow}
\usepackage{epstopdf}
\usepackage{longtable}
\usepackage{graphicx,color,dcolumn,booktabs,bm}
\usepackage[colorlinks,
            citecolor=blue,
            anchorcolor=red,
            menucolor=red,
            linkcolor=red,
            filecolor=red,
            runcolor=red,
            urlcolor=blue,
            frenchlinks=red]{hyperref}

\graphicspath{{Figures/}} %

\allowdisplaybreaks

\begin{document}

\title{Probing double hadron resonances by the complex scaling method}

\author{Zhuo Yu$^{2}$}
\author{Mao Song$^{1}$}
\email{songmao@mail.ustc.edu.cn}
\author{Jian-You Guo$^{1}$}
\author{Yu Zhang$^{1,2}$}
\author{Gang Li$^{1}$}

\affiliation{ $^1$School of Physics and Material Science, Anhui University, Anhui University, Hefei 230601, China
\\
$^2$Institutes of Physical Science and Information Technology, Anhui University, Hefei 230601, China}

\begin{abstract}
Many newly discovered excited states are interpreted as bound states of hadrons. Can these hadrons also form resonant states?
In this paper, we extend the complex scaling method (CSM) to calculate the bound state and resonant state consistently for the $\Lambda_c D(\bar D)$ and $\Lambda_c \Lambda_c (\bar \Lambda_c)$ systems. For these systems, the $\pi, \eta, \rho$ meson exchange contributions are suppressed, the contributions of intermediate- and short-range forces from $\sigma/\omega$ exchange are dominant. Our results indicate that $\Lambda_c D$ system can not form bound state and resonant state. There exist resonant states in a wide range of parameters for $\Lambda_c \bar D$ and $\Lambda_c \Lambda_c (\bar \Lambda_c)$ systems. For these systems, the larger bound state energy, the easier to form resonant states. Among all the resonant states, the energies and widths of the P wave resonant states are smaller and more stable, which is possible to be observed in the experiments. The energies of D and F wave resonant states can reach dozens of MeV and the widths can reach hundreds of MeV.
\end{abstract}


\maketitle

\section{introduction}\label{sec1}

Since the observations of the charmonium-like state $X(3872)$ \cite{Choi:2003ue}, many new type hadrons have been discovered. The quantum numbers of these states are different from the traditional $q\bar{q}$ mesons and $qqq$ baryons, so they are called the exotic hadrons, such as $X(3872)$ \cite{Choi:2003ue}, $Y(4260)$/$Y(4360)$ \cite{Aubert:2005rm, Aubert:2006ge}, $Z_b(10610)/Z_b(10650)$ \cite{Belle:2011aa}, $P_c(4380)/P_c(4450)$ \cite{Aaij:2015tga}. Since many of these exotic states $X/Y/Z/P_c$ are close to the thresholds of two hadrons, they are naturally considered as candidates of the hadronic molecules. The non-relativistic effective field theories and lattice QCD are the suitable to study the structure of these hadronic molecules.

Among various explanations, hadronic molecules gain more attention, especially for the hadrons containing one or two heavy quark(s). Deuteron is the only stable molecule state of hadrons, which can be well explained by the one-boson-exchange model \cite{Tornqvist:1993vu,Tornqvist:1993ng}. Along this way, we suppose that the deuteron-like hadron states have similar structures and interaction potentials. The one-boson-exchange model is a reasonable theoretical method for explaining the hadronic molecular states. According to the mass of exchanged mesons, the mesons $\pi$, $\eta$, $\sigma$, $\rho$ and $\omega$ exchange contribute to the long, medium, and short range interaction, respectively. In the last decade, the molecular states of hadrons have been extensively explored for the discoveried charmonium-like $X/Y/Z$ and $P_c(4380)/P_c(4450)$ states \cite{Liu:2008fh,Thomas:2008ja,Lee:2009hy,Liu:2009ei,Sun:2011uh,Chen:2015loa}. In the framework of one-boson-exchange model, these hadrons can not only bind to hadron molecular states, but also form resonant states with high angular momentum, which is less studied in hadron physics.

The resonance is one of most striking phenomenon in the scattering experiment, which exists widely in atoms, molecules, nuclei and chemical reactions. Therefore, researchers have developed many methods to study resonances, including: $R$-matrix method \cite{Wigner:1947zz, Hale:1987zz}, $K$-matrix method \cite{Humblet:1991zz}, scattering phase shift method, continuous spectrum theory, $J$-matrix method \cite{Taylor}, coupling channel method, real stabilization method (RSM)\cite{Hazi}, analytic continuation method of coupling constant (ACCC) \cite{Kukulin} and complex scaling method (CSM) \cite{csm1,csm2}, $etc$. Among them, RSM, ACCC and CSM are bound-state-type methods, which can be conveniently dealt with bound-like state. In the framework of non-relativistic theory, these methods have obtained dramatically improvements. For example, RSM has been able to effectively determine the parameters of the resonant state through improved calculation methods. In combination with the cluster model, the ACCC has been used to calculate the energy and lifetime of some light nuclear resonant states, as well as the wave function of the resonant states. The complex scaling method can describe the bound state, resonant state and continuum in a consistent way, which is widely used to exploring the resonance in atomic, molecular and nuclear physics. The CSM have been extended from nonrelativistic to the relativistic framework \cite{csm3,csm4,csm5,csm6,csm7}and from spherical nuclei to deformed nuclei \cite{csm8,csm9}, which have been applied in halo nuclei.

We will apply CSM to hardon physics for searching the resonant states. In the system of heavy molecular states, we need consider the interaction of $\pi$, $\eta$, $\rho$, $\sigma$, $\omega$ mesons. Unusually, the effect of one-$\pi$ meson exchange is dominant, and obscures the contribution of other mesons. In Ref. \cite{Chen:2017jjn}, the authors have discussed the bound states of $DD(\bar{D})$, $\Lambda_cD(\bar{D})$ and $\Lambda_c\Lambda_c(\bar{\Lambda}_c)$ through one-$\sigma$-exchange(OSE) and one-$\omega$-exchange(OOE). Based on the spin and isospin conservation, there is no coupling $\pi \Lambda_c \Lambda_c$ and $\pi D D$, and the contribution of $\pi$, $\eta$, $\rho$ meson exchanges are forbidden or suppressed heavily. The $\Lambda_c\bar{\Lambda}_c$ system has been investigated in several previous works \cite{Meguro:2011nr,Li:2012bt,Lee:2011rka}. In this paper, we will further investigate whether two hardons are possible to form resonant states by exchanging $\sigma$ and $\omega$ mesons for $DD(\bar{D})$, $\Lambda_cD(\bar{D})$ and $\Lambda_c\Lambda_c(\bar{\Lambda}_c)$ systems.

This paper is organized as follows. After the introduction, we present the theoretical framework and calculation method in Section \ref{sec2}. The numerical results and discussion are given in Section \ref{sec3}. A short summary is given in Section \ref{sec4}.

\section{Theoretical framework}\label{sec2}

Under the heavy-quark symmetry, the effective Lagrangians for one-$\sigma$-exchange and one-$\omega$-exchange are expressed,
\begin{eqnarray}
\mathcal{L}_{DD\sigma/\omega} &=& -2g_{\sigma}D^{\dag}D\sigma+2g_{\omega} D^{\dag}D \upsilon \cdot \omega,\label{DD}\\
\mathcal{L}_{\Lambda_c\Lambda_c\sigma/\omega} &=& -2g_{\sigma}'\bar{\Lambda}_c\Lambda_c\sigma-2g_{\omega}'\bar{\Lambda}_c\Lambda_c \upsilon \cdot \omega.\label{LL}
\end{eqnarray}
Here, $\upsilon$ is the four velocity of the heavy meson, which has the form of $\upsilon=(1,\bf{0})$.

The coupling strengths can be estimated by using the quark model, where the $\sigma$ and $\omega$ mesons couple to light quarks in heavy hadrons.
Since the $\sigma$ and $\omega$ mesons couple dominantly to the light quarks, the Lagrangian of the light quarks $(q=u,d)$ and $\sigma/\omega$ can be written as
\begin{eqnarray}
\mathcal{L}_{qq\sigma/\omega} &=& -g_{\sigma}^{q}\bar{\psi}_q\sigma\psi_q-g_{\omega}^{q}\bar{\psi}_q\gamma^{\mu}\omega_{\mu}\psi_q.\label{quark}
\end{eqnarray}
Compared with the vertices of $D^{\dag}D\sigma/\omega$, $\bar{\Lambda_c}\Lambda_c\sigma/\omega$, and $\bar{q}q\sigma/\omega$ in Eqs. (\ref{DD})$-$(\ref{quark}), the coupling constants can be related, i.e.,
\begin{eqnarray}\label{cc}
g_{\sigma} = g_{\sigma}'=g_{\sigma}^{q}, \quad\quad
g_{\omega} = g_{\omega}'=g_{\omega}^{q}.
\end{eqnarray}
In a $\sigma$ model \cite{Riska:1999fn}, the value of $g_{\sigma}^{q}$ is taken as $g_{\sigma}^{q}=3.65$. For the $\omega$ coupling $g_{\omega}^{q}$, in the Nijmegen model, $g_{\omega}^{q}=3.45$, whereas it is equal to 5.28 in the Bonn model \cite{Rijken:1998yy}. In Ref. \cite{Riska:2000gd}, $g_{\omega}^{q}$ was roughly assumed to be 3.00. In the following calculation, all the possible choices will be considered.

According to the effective Lagrangians in Eqs. (\ref{DD}) and (\ref{LL}), all the relevant OBE scattering amplitudes can be collected in Table \ref{amplitude}.

\renewcommand\tabcolsep{0.1cm}
\renewcommand{\arraystretch}{2.2}
\begin{table}[htbp]
  \caption{The scattering amplitudes for all investigated systems. Here, $\mathcal{H}(\bm{q},m)$ is defined as $\mathcal{H}(\bm{q},m)=1/(\bm{q}^2+m^2)$.}\label{amplitude}
  \begin{tabular}{cl}\toprule[2pt]
  $h_1h_2\to h_3h_4$            &$\mathcal{M}(h_1h_2\to h_3h_4)$ \\\hline
  $DD\to DD$
                   &$4M_D^2\left[g_{\sigma}^2\mathcal{H}(\bm{q},m_{\sigma})-g_{\omega}^2\mathcal{H}(\bm{q},m_{\omega})\right]$\\
  $D\bar{D}\to D\bar{D}$
                   &$4M_D^2\left[g_{\sigma}^2\mathcal{H}(\bm{q},m_{\sigma})+g_{\omega}^2\mathcal{H}(\bm{q},m_{\omega})\right]$\\
  $\Lambda_c\bar{D}\to \Lambda_c\bar{D}$
                   &$8M_{D}M_{\Lambda_c}\chi_3^{\dag}\chi_1\left[g_{\sigma}g_{\sigma}'\mathcal{H}(\bm{q},m_{\sigma})-g_{\omega}g_{\omega}'\mathcal{H}(\bm{q},m_{\omega})\right]$\\
  $\Lambda_c{D}\to \Lambda_c{D}$
                   &$8M_{D}M_{\Lambda_c}\chi_3^{\dag}\chi_1\left[g_{\sigma}g_{\sigma}'\mathcal{H}(\bm{q},m_{\sigma})+g_{\omega}g_{\omega}'\mathcal{H}(\bm{q},m_{\omega})\right]$\\
  $\Lambda_c\Lambda_c\to\Lambda_c\Lambda_c$
       &$16M_{\Lambda_c}^2\chi_3^{\dag}\chi_4^{\dag}\chi_1\chi_2\left[g_{\sigma}g_{\sigma}'\mathcal{H}(\bm{q},m_{\sigma})-g_{\omega}g_{\omega}'\mathcal{H}(\bm{q},m_{\omega})\right]$\\
  $\Lambda_c\bar{\Lambda}_c\to\Lambda_c\bar{\Lambda}_c$
       &$16M_{\Lambda_c}^2\chi_3^{\dag}\chi_4^{\dag}\chi_1\chi_2\left[g_{\sigma}g_{\sigma}'\mathcal{H}(\bm{q},m_{\sigma})+g_{\omega}g_{\omega}'\mathcal{H}(\bm{q},m_{\omega})\right]$\\
  \bottomrule[2pt]
  \end{tabular}
\end{table}

According to the G-parity rule \cite{Klempt:2002ap}, the OSE and OOE effective potentials for the $D\bar{D}$, $\Lambda_c\bar{D}$, and $\Lambda_c\bar{\Lambda}_c$ systems are related to the potentials for the $D D$, $\Lambda_c D$, and $\Lambda_c \Lambda_c$ systems. The interactions from the $\sigma$ exchange are same, and from the $\omega$ exchange are contrary for $D\bar{D}$ and $D D$, $\Lambda_c\bar{D}$ and $\Lambda_c D$, $\Lambda_c\bar{\Lambda}_c$ and $\Lambda_c \Lambda_c$ systems.

With the Breit approximation, one can get the relation between the effective potentials in momentum space $\mathcal{V}_{fi}$ and the scattering amplitude $\mathcal{M}_{fi}$ in the momentum space, i.e.,
\begin{eqnarray}
\mathcal{V}_{fi}(\bm{q}) &=& -\frac{\mathcal{M}_{fi}(h_1h_2\to h_3h_4)}{\sqrt{\prod_i2m_i\prod_f2m_f}}.
\end{eqnarray}
Here, $\mathcal{M}(h_1h_2\to h_3h_4)$ is the scattering amplitude for the process $h_1h_2\to h_3h_4$. $m_i$ and $m_f$ are the masses of the initial ($h_1$, $h_2$) and final particles ($h_3$, $h_4$), respectively.

The effective potential in the coordinate space $\mathcal{V}(r)$ is obtained by performing the Fourier transformation as
\begin{eqnarray}\label{fourier}
\mathcal{V}(\bm{r}) &=& \int\frac{d^3\bm{q}}{(2\pi)^3}e^{i\bm{q}\cdot\bm{r}}\mathcal{V}(\bm{q})\mathcal{F}^2(q^2).
\end{eqnarray}

In order to regularize the off shell effect of the exchanged mesons and the structure effect of the hadrons, a monopole form factor $\mathcal{F}(q^2)$ is introduced at every vertex,

\begin{eqnarray}
\mathcal{F}(q^2) &=& \frac{\Lambda^2-m^2}{\Lambda^2-q^2},
\end{eqnarray}
here, $\Lambda$ is the cutoff parameter, $m$ and $q$ correspond to the mass and momentum of the exchanged meson, respectively.
In Refs. \cite{Tornqvist:1993vu,Tornqvist:1993ng}, $\Lambda$ is related to the root-mean-square radius of the source hadron which propagate the interaction through the intermediate boson ($\sigma$ or $\omega$). According to the previous experience of the deuteron, the cutoff $\Lambda$ is taken around 1.0 GeV.

After adding the monopole form factor $\mathcal{F}(q^2)$, the effective potentials in coordinate space are obtained in Table \ref{tot}.
\renewcommand\tabcolsep{0.2cm}
\renewcommand{\arraystretch}{1.8}
\begin{table}[!htbp]
  \caption{The effective potentials in coordinate space for all investigated systems. The function $Y(\Lambda,m,r)$ is defined as $Y(\Lambda,m,{r}) = (e^{-mr}-e^{-\Lambda r})/4\pi r-(\Lambda^2-m^2)e^{-\Lambda r}/{8\pi \Lambda}$.}\label{tot}
  \begin{tabular}{ccl}\toprule[2pt]
  Systems      &Quarks    &$\mathcal{V}({r})$ \\\hline
  $DD$          &$(c\bar{q})(c\bar{q})$           &$-g_{\sigma}^2Y(\Lambda,m_{\sigma},r)+g_{\omega}^2Y(\Lambda,m_{\omega},r)$\\
     $D\bar{D}$          &$(c\bar{q})(\bar{c}{q})$           &$-g_{\sigma}^2Y(\Lambda,m_{\sigma},r)-g_{\omega}^2Y(\Lambda,m_{\omega},r)$\\
  $\Lambda_c\bar{D}$     &$(cqq)(\bar{c}{q})$         &$-2g_{\sigma}g_{\sigma}'Y(\Lambda,m_{\sigma},r)+2g_{\omega}g_{\omega}'Y(\Lambda,m_{\omega},r)$\\
     $\Lambda_c{D}$     &$(cqq)(c\bar{q})$         &$-2g_{\sigma}g_{\sigma}'Y(\Lambda,m_{\sigma},r)-2g_{\omega}g_{\omega}'Y(\Lambda,m_{\omega},r)$\\
  $\Lambda_c\Lambda_c$    &$(cqq)(cqq)$          &$-4g_{\sigma}^{'2}Y(\Lambda,m_{\sigma},r)+4g_{\omega}^{'2}Y(\Lambda,m_{\omega},r)$\\
      $\Lambda_c\bar{\Lambda}_c$    &$(cqq)(\bar{c}\bar{q}\bar{q})$          &$-4g_{\sigma}^{'2}Y(\Lambda,m_{\sigma},r)-4g_{\omega}^{'2}Y(\Lambda,m_{\omega},r)$\\
  \bottomrule[2pt]
  \end{tabular}
\end{table}

In Table \ref{tot}, we can find that the interactions of one-$\sigma$-exchange are always attractive for these systems. The depth of the one-$\sigma$-exchange effective potentials depend on the number of the light quarks and/or antiquark combinations $(qq, q\bar{q}, \bar{q}\bar{q})$, where the light quark or anti-quark is reserved in different hadrons of the hadron-hadron systems, respectively.

The one-$\sigma$-exchange and one-$\omega$-exchange interactions are corresponding intermediate- and short-range forces, and therefore they are suppressed when the radius $r$ reaches 1.0 fm or larger. Since the force of the one-$\sigma$-exchange is the dominant, the total effective potentials for all the  systems are attractive. The one-$\sigma$-exchange can always provide an attractive force. However, the one-$\omega$-exchange is repulsive for the system including the same light quarks or antiquarks in its components of the investigated systems. The interaction strengths for the one-$\sigma$-exchange and one-$\omega$-exchange depend on the light-quark combination numbers.

We extend the CSM to describe the resonance of $\Lambda_c D(\bar D)$ and $\Lambda_c \Lambda_c(\bar \Lambda_c)$ systems. The advantage of this approach is that both bound and resonant states can be treated consistently, since the complex scaled functions of the resonant states are square integrable same as the bound state. The Hamiltonian can be written as:
\begin{equation}
H=T+V\ ,  \label{Hamiltonian}
\end{equation}%
where $T$ is the kinetic energy, and $V$ is the effective potential of the system, which are obtained from the double hadron scattering. In the CSM, the relative coordinate $r$ in Hamiltonian $H$ and wave function $\psi$ is complex scaled as
\begin{equation}
U(\theta ):r\rightarrow re^{i\theta }.  \label{tranformation}
\end{equation}%

We can get the transformed Hamiltonian and the wave function $H_{\theta }=U\left( \theta \right) HU^{-1}\left( \theta \right) $ and $\psi
_{\theta }=U\left( \theta \right) \psi $, where $\psi _{\theta }$ is square integrable. Then, the corresponding complex scaled equation is obtained,

\begin{equation}
H_{\theta }\psi _{\theta }=E_{\theta }\psi _{\theta }.  \label{Complexeq}
\end{equation}%
Based on the Aguilar-Balslev-Combes(ABC) theorem\cite{abc}, the energy spectrum is a set of poles of the Green function in the complex
energy plane, which consists of three parts: (i) the bound states
are discrete set of real points on the negative energy axis (ii) the resonant states correspond to the discrete set
of points in the lower half of complex energy plane; and (iii) the continuous spectrum is rotated around the origin of the complex energy plane by an
angle $2\theta$.

To solve the complex scaled equation, the basis expansion method is applied. The total wave function $\psi _{\theta }$ can be expanded as
\begin{equation}
\psi _{\theta }=\sum_{i}c_{i}\left( \theta \right) \phi _{i}\text{ },
\label{Expand}
\end{equation}%
where $\phi _{i}=R_{nl}(r)Y_{lm_{l}}(\vartheta ,\varphi )$ and the index $i$ sum over all the quantum numbers $n,l,m_{l}$. $R_{nl}(r)$ is the radial function of a spherical harmonic oscillator potential,
\begin{equation}
R_{nl}(r)=\frac{1}{b_{0}^{3/2}}\sqrt{\frac{2\left( n-1\right) !}{\Gamma
\left( n+l+1/2\right) }}x^{l}L_{n-1}^{l+1/2}\left( x^{2}\right) e^{-x^{2}/2}%
\text{, }n=1,2,3,\text{ }....  \label{hobasis}
\end{equation}%
Here, $x=r/b_{0}$ is the radius measured in units of the oscillator length $b_{0}$.
$Y_{lm_{l}}(\vartheta ,\varphi )$ is the spherical harmonics, and describes the angular distribution of particles.
Inserting the wave function (\ref{Expand}) into the complex scaled equation (\ref{Complexeq}) and applying the
orthogonality of wave functions $\phi _{i}$, we get a symmetric
matrix diagonalization,
\begin{equation}
\sum_{i}\left[ T_{i^{\prime },i}+V_{i^{\prime },i}\right] c_{i}=E_{\theta
}c_{i^{\prime }}\text{ },  \label{matrixeq}
\end{equation}%
where $T_{i^{\prime },i}$ and $V_{i^{\prime },i}$ are presented as%
\begin{eqnarray}
T_{i^{\prime },i} &=&e^{-i2\theta }\int \phi _{i^{\prime }}\left( -\frac{%
\hbar ^{2}}{2M}\left( \frac{d^{2}}{dr^{2}}+\frac{2}{r}\frac{d}{dr}\right) +%
\frac{\vec{l}^{2}}{2Mr^{2}}\right) \phi _{i}d\vec{r}\text{ },
\label{Tmatrix} \nonumber \\
V_{i^{\prime },i} &=&\int \phi _{i^{\prime }}V\left( \vec{r}e^{i\theta
}\right) \phi _{i}d\vec{r}\text{ }.  \label{Vmatrix}
\end{eqnarray}%
Here, $M=m_1m_2/(m_1+m_2)$ is the reduced mass of particle 1 and particle 2 system. Substituting $\phi _{i}$ into the above equation, the matrix elements $T_{i^{\prime },i}$ are obtained as

\begin{eqnarray}
T_{i^{\prime },i} &=& e^{-i2\theta }\frac{\hbar^{2}}{2Mb_{0}^{2}} [\sqrt{n \left( n+l+1/2\right)}\delta_{n^{\prime},n+1} \nonumber \\
&+&(2n+l-1/2)\delta_{n^{\prime },n}  \nonumber \\
&+& \sqrt{(n-1) \left( n+l-1/2 \right) }\delta_{n^{\prime},n-1} ]
\delta_{l^{\prime }l}\delta_{m_{l}^{\prime}m_{l}}.  \label{Telement}
\end{eqnarray}

Similarly, the matrix elements $V_{i^{\prime },i}$ are obtained as
\begin{eqnarray}
V_{i^{\prime },i} &=&\left\langle n^{\prime }l^{\prime }\right\vert
V(re^{i\theta })\left\vert nl\right\rangle \delta _{l^{\prime }l}\delta
_{m_{l}^{\prime }m_{l}},  \nonumber \\
 &=& \sum_{n=1}^N \Lambda_{nk}\Lambda_{n^{\prime}k} [\epsilon_k U(\epsilon_k/\omega)] \label{Vcent}
\end{eqnarray}%
$\epsilon_k$ and $\Lambda_{nk}$ are the eigenvalues and eigenvectors of the matrix $JK$ with elements
$JK_{n,n} = 2 (n + l)$; $JK_{n,n+1} = - \sqrt{n (n + 2l + 1)}$.
With the matrix elements $T_{i^{\prime },i}$ and $V_{i^{\prime },i}$, the
solutions of the complex scaled equation (\ref{Complexeq}) can be obtained
by diagonalizing the matrix $H_{\theta }$. The eigenvalues of $H_{\theta }$
representing bound states or resonant states do not change with $\theta $,
while the eigenvalues representing the continuous spectrum rotate with $%
\theta $. The former are associated with resonance complex energies $%
E-i\Gamma /2$, where $E$ is the resonance position and $\Gamma $ is its
width.

\section{Numerical results}\label{sec3}

\begin{figure}
\centering
\includegraphics[width=3.33in, keepaspectratio]{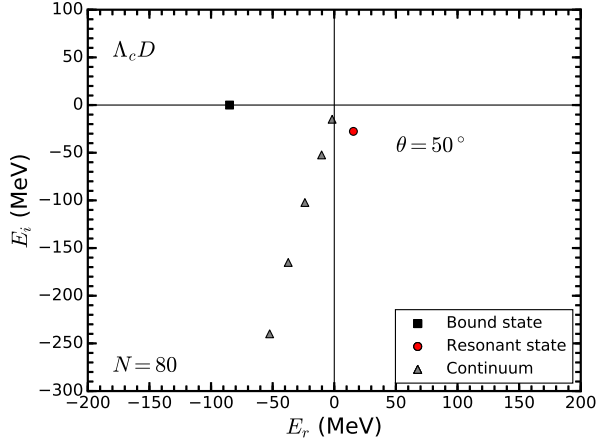}\\
\caption{(Color online) The resonant state is presented with $\theta = 50^{\circ}$. Here, the cutoff parameter $\Lambda=1.2$ GeV, and the $\omega$ coupling constant $g_\omega=5.28$. The result is performed by expending the basis function with N=80.}\label{fig1}
\end{figure}

\begin{figure}
\centering
\includegraphics[width=3.33in, keepaspectratio]{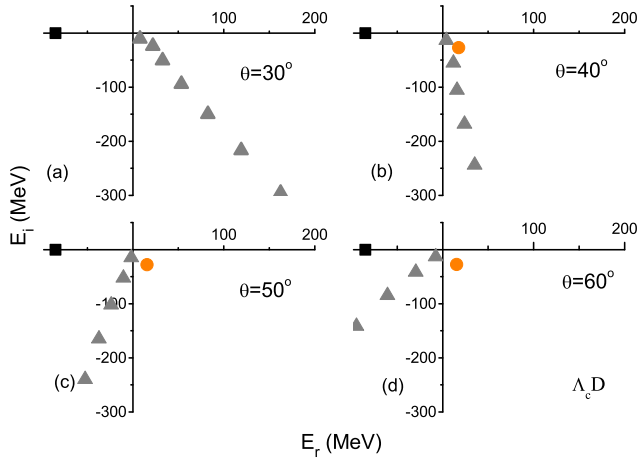}\\
\caption{(Color online) The resonant and continuous spectra varying with the complex rotation angle in the complex energy plane.
Except for the complex rotation angle, the other parameters are set same as those in Fig. \ref{fig1}.}\label{fig2}
\end{figure}

Our purpose is to extend the CSM to describe the resonances of two hardons. In this section, we discuss and analysis the effects of the OSE and OOE interactions for the systems of $\Lambda_c D(\bar D)$ and $\Lambda_c \Lambda_c (\bar \Lambda_c)$ by solving the Schr\"{o}dinger equation in the CSM. The related parameters are summarized in Table \ref{parameter}.
\renewcommand\tabcolsep{0.16cm}
\renewcommand{\arraystretch}{1.8}
\begin{table}[!htbp]
  \caption{The related parameters are used in this work \cite{Olive:2016xmw}.}\label{parameter}
  \begin{tabular}{ccc|ccc}\toprule[2pt]
  Hadron     &$I(J^P)$     &Mass (MeV)    &Hadron     &$I(J^P)$     &Mass (MeV) \\\hline
  $D$        &$\frac{1}{2}(0^-)$    &1867.24      &$\Lambda_c$     &$0(\frac{1}{2}^+)$     &2286.46\\
  $\sigma$   &$0(0^+)$              &600          &$\omega$        &$0(1^-)$               &782.65\\
  \bottomrule[2pt]
  \end{tabular}
\end{table}
 The information about the resonant state is obtained after we diagonalize the Hamiltonian. We take $\Lambda$-D system as an example to describe how to find out the resonant state by complex rotation. The process is shown in Figs. \ref{fig1} and \ref{fig2}.

As shown in Fig. \ref{fig1}, all the eigenvalues of $H_\theta$ are divided into three parts: the bound state, the resonant state, and the continuum, respectively, labeled as black squares, orange solid dots and gray triangles. The bound state locates on the negative energy axis, while the
continuous spectrum of $H_\theta$ rotates clockwise with the angle $2\theta$ and resonant state locates in the lower half of the complex energy
plane, which is bounded by the rotated continuum line and the positive energy axis and become isolated.

In order to clarify how the resonant states isolate from continuum by complex rotation, we present the variation of resonant states with the complex scale angle $\theta$, and the other parameters are fixed same as in Fig. \ref{fig1}. In Fig. \ref{fig2}(a), when $\theta = 30^{\circ}$, all dots are nearly in one straight line, and it is hard to distinguish resonant state in the continuous spectrum. In Fig. \ref{fig2}(b), when $\theta = 40^{\circ}$, the resonant state begin to separate from the continuous spectrum. When complex scale angle $\theta$ reaches $50^{\circ}$ in Fig. \ref{fig2}(c), the resonant state is completely separated from the continuous spectrum. Then, with the angle $\theta$ increasing, the continuum spectra rotate with, the position of the resonant state in the complex plane is almost not movement as shown in Fig. \ref{fig2}(d).

\begin{figure}
\centering
\includegraphics[width=3.33in, keepaspectratio]{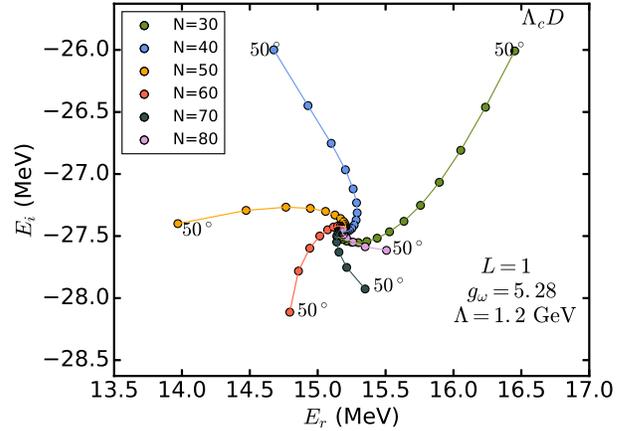}\\
\caption{(Color online) The $\theta$ trajectories corresponding to the
several different numbers of oscillator shells of the basis for the
resonant state, where N is the quantum number of the main shell of the oscillator basis and $\theta$
varies from $50^{\circ}$ to $85^{\circ}$ by steps of $2.5^{\circ}$. }\label{fig3}
\end{figure}
In the complex scaling method, the resonance energy should not depend on the choice of angle $\theta$. In practice, the number of basis chosen is finite for basis expansion method, the position of resonance state is not completely independent on $\theta$. In order to get a reasonable value for the number of basis N, we plot the trajectories of the resonant energies for different values of N with angle $\theta$ in Fig. \ref{fig3}. We can see that as the number of basis increases from 30 to 80, and the angle $\theta$ increase from $50^{\circ}$ to $85^{\circ}$, these trajectories tend to converge to same location. This means that the numerical results are insensitive to the number of basis N, and for any number N between 30 to 80, we can obtain the sufficient precision numerical value.

\begin{figure}
\centering
\includegraphics[width=3.33in, keepaspectratio]{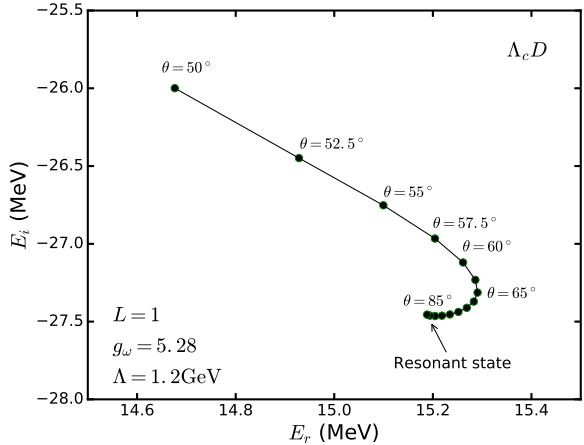}\\
\caption{(Color online) The parameters are same in Fig. \ref{fig3}, but a closeup of the $\theta$ trajectory in
Fig. 4 for N = 80, where the arrow point to the position of the resonant state.}\label{fig4}
\end{figure}

In each $\theta$ trajectory, there is a point corresponding to the minimal rate of change with the rotation angle $\theta$,
which indicate the optimal value for the resonance parameters. From the Fig. \ref{fig3}, we know that the resonant state is a function of $\theta$, and the existence of minimal rate of change point indicates that the energy of the resonant state has the optimal value for the rate of change $\theta$. In order to better determine this point, we show the $\theta$ trajectory of N=80 in Fig. \ref{fig4}, with the other parameters are same as in Fig. \ref{fig3}. The optimal energy value of the resonant state appears at around $\theta = 85^{\circ}$, which is almost independent of $\theta$, i.e.$\frac{dE}{d\theta}=0$.

\renewcommand\tabcolsep{0.3cm}
\renewcommand{\arraystretch}{1.8}
\begin{table*}[!htbp]
  \caption{ The energy and width of bound and resonant states for the $\Lambda_cD(\bar{D})$ and $\Lambda_c\Lambda_c(\bar{\Lambda}_c)$ systems. $E_r$ and $\Gamma$ represent the energy and width of resonant states in units of MeV, respectively. The cutoff $\Lambda$ is set as 1.1GeV. The value of coupling constant $g_{\omega}^{q}$ are set as  $g_{\omega}^{q}=3.00$ in Ref. \cite{Riska:2000gd}, 3.45 in the Nijmegen model, and 5.28 in the Bonn model \cite{Rijken:1998yy}, respectively. The notation $\ldots$ stands for no bound or resonant state solutions.}\label{num2}
  \begin{tabular}{c|ccccccc||ccccccc}\toprule[2pt]
     \multirow{2}*{\,$g_{\omega}^{q}$\,}     &\multicolumn{3}{c}{$\Lambda_c\bar D$}    &    &\multicolumn{3}{c||}{$\Lambda_cD$}     &\multicolumn{3}{c}{$\Lambda_c\Lambda_c$}    &   &\multicolumn{3}{c}{$\Lambda_c\bar{\Lambda}_c$}\\\cline{2-15}

                        &L   &$E$      &$\Gamma$       &           &L   &$E$      &$\Gamma$
                        &L   &$E$      &$\Gamma$       &           &L   &$E$      &$\Gamma$\\\hline
  3.0                   &S   &\ldots      &\ldots         &           &S    &-11.09         &\ldots
                        &S   &-31.69      &\ldots         &           &S    &-110.39         &\ldots\\

                        &P   &\ldots      &\ldots         &           &P    &\ldots         &\ldots
                        &P   &7.74        &55.57    &           &P    &10.30        &19.27\\

                        &D   &\ldots      &\ldots         &           &D    &\ldots         &\ldots
                        &D   &\ldots      &\ldots         &          &D    &28.67         &196.22\\

                        &F   &\ldots      &\ldots         &           &F    &\ldots         &\ldots
                        &F   &\ldots      &\ldots         &           &F    &2.06         &438.05\\\hline

  3.45                  &S   &\ldots      &\ldots         &           &S    &-14.40        &\ldots
                        &S   &-22.74       &\ldots         &           &S    &-126.02        &\ldots\\

                        &P   &\ldots      &\ldots         &           &P    &0.27           &93.03
                        &P   &5.58        &60.43          &           &P    &9.07           &14.09\\

                        &D   &\ldots      &\ldots         &           &D    &\ldots         &\ldots
                        &D   &\ldots      &\ldots         &           &D    &32.93        &190.20\\

                        &F   &\ldots      &\ldots         &           &F    &\ldots         &\ldots
                        &F   &\ldots      &\ldots         &           &F    &10.81         &437.80\\\hline

  5.28                  &S               &       &\ldots  &           &S    & -38.66        &\ldots
                        &S       &\ldots       &\ldots      &          &S           &-222.57       &\ldots \\

                        &P       &\ldots        &\ldots      &    &P             &9.64        &75.53
                        &P       &\ldots        &\ldots             &    &P             &\ldots        &\ldots \\

                        &D       &\ldots        &\ldots      &    &D        &\ldots        &\ldots
                        &D       &\ldots        &\ldots             &    &D        &49.22       &149.69\\

                        &F   &\ldots      &\ldots         &           &F    &\ldots         &\ldots
                        &F   &\ldots      &\ldots         &           &F    &52.46         &417.68\\
  \bottomrule[2pt]
  \end{tabular}
\end{table*}

In Table \ref{num2}, we present the numerical results of bound and resonant states for the  $\Lambda_cD(\bar{D})$ and $\Lambda_c\Lambda_c(\bar{\Lambda}_c)$ systems with $\Lambda=1.1$ GeV. By solving the Schr\"{o}dinger equation with CSM, the solutions are divided into two parts: one is the bound state solution with negative energy; others are resonant states, and have the form $E-i\Gamma /2$, where $E$ is the resonance energy and $\Gamma $ is its width. In the table, we list the energies of the bound states, the energies and widths of the resonant states for different angular momentum $L$. We find that there are neither bound states nor resonant states for $DD$ and $D\bar{D}$ systems. The total interaction of the $\Lambda_c\bar D$ system includes the $\sigma$ exchange attraction and the $\omega$ exchange repulsion. As the cutoff parameter $\Lambda$ is increased, the $\sigma$ meson exchange becomes more prominent. Due to the stronger $\omega$ exchange repulsion, they can not form bound state, much less to resonant states.

Compared to the $\Lambda_c\bar D$ system, both the $\sigma$ and $\omega$ meson provide attractive contribution in the $\Lambda_c D$ system, and it is strong enough to form a shallow S wave bound state. For cases $g_{\omega}^{q}=3.45$ and 5.28, this system can form a P wave resonant state, the energy is about several MeV, the width arrive dozens of MeV.

For the $\Lambda_c\Lambda_c(\bar{\Lambda}_c)$ systems, the interaction strength is two times stronger than
that in the $\Lambda_c\bar{D}$ and $\Lambda_cD$ systems. Therefore, it is more easy form bound and resonant states for $\Lambda_c\Lambda_c$ and $\Lambda_c\bar{\Lambda}_c$ systems. When the cutoff $\Lambda$ = 1.1 GeV, the binding energies can reach from several to dozens MeV. For $\Lambda_c\Lambda_c$ system, they can form a P wave resonant state; For $\Lambda_c\bar{\Lambda}_c$ system, they can form more high angular momentum resonant states.

\begin{figure}
\centering
\includegraphics[width=3.33in, keepaspectratio]{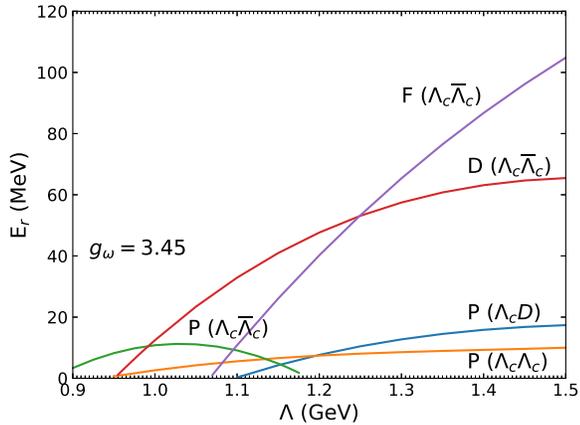}\\
\caption{(Color online) The energy of the resonant state as a function of the cutoff parameter $\Lambda$ for these systems
with $g_{\omega}^{q}=3.45$. }\label{fig5}
\end{figure}

\begin{figure}
\centering
\includegraphics[width=3.33in, keepaspectratio]{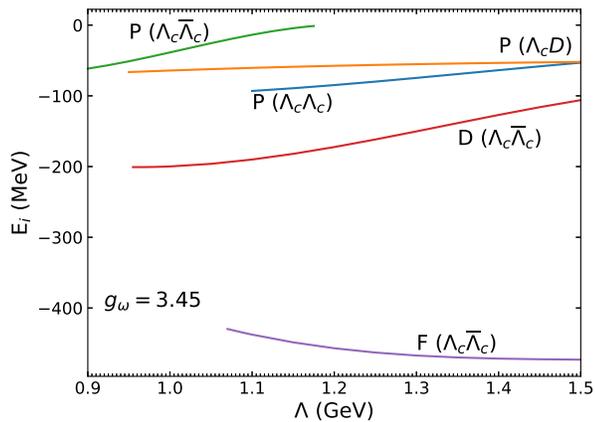}\\
\caption{(Color online) The width of the resonant state as a function of the cutoff parameter $\Lambda$ for these systems with $g_{\omega}^{q}=3.45$. }\label{fig6}
\end{figure}

The cutoff parameter $\Lambda$ has a great effect on the energy and width of the resonant states. In Fig. \ref{fig5} and Fig. \ref{fig6}, we present the energy and width curves as a function of $\Lambda$. From Fig. \ref{fig5} and Fig. \ref{fig6}, we can see that the energy of the P wave resonant state is only a few to a dozen MeV, and the corresponding width is dozens of MeV with $\Lambda$ changing from 0.9 to 1.5 GeV. With the variation of $\Lambda$, the change of energy and width was not very dramatic. For the D and F wave resonant states of $\Lambda_c\bar{\Lambda}_c$ system, with the increasing of $\Lambda$, the resonant energy can reach dozens or even a hundred MeV, the corresponding widths can reach about 200 MeV and 480 MeV. The resonance width increases with the increasing of
angular momentum L for $\Lambda_c\bar{\Lambda}_c$ system, which means the D and F wave resonant states become even more unstable.

\section{Summary}\label{sec4}
A large number of new excited hadronic states have been found in experiments, and many excited states
are explained by the hadronic molecular states. We extend the complex scaling method(CSM) to
describe bound and resonant states. For $\Lambda_cD (\bar{D})$ and  $\Lambda_c \Lambda_c (\bar \Lambda_c)$ systems, under the heavy quark symmetry,
the contributions of $\pi, \eta, \rho$ meson exchanges are greatly suppressed,
and the contribution of the $\sigma$ and $\omega$ mesons exchange are dominant in these process.
Our results indicate that there may be exist resonant states for $\Lambda_cD (\bar{D})$ and  $\Lambda_c \Lambda_c (\bar \Lambda_c)$ systems.
The resonant states position in the complex energy plane remains almost unchanged with the variation of $\theta $.
We also checks the independent of basis numbers N with $\theta$ trajectories, the results show that the numerical precision is
insensitive to the basis numbers. Our results indicate that there may be form resonant states for $\Lambda_cD$ and $\Lambda_c \Lambda_c (\bar \Lambda_c)$ systems. In these resonant states, the P waves have relative smaller energies and widths, which is more stable and easy to search in the experiments.
In general, for these systems, the resonant states are possible to form in a wide range of parameters, and may be observed in experiments in the future.


\vfill

\section*{ACKNOWLEDGMENTS}
The author thanks the useful discussions with Kan Chen, Ran Ding and Quan Liu. This work was supported in part by the National Natural Science Foundation of China (No.11805001, No.11935001).


\begin{thebibliography}{99}

\bibitem{Choi:2003ue}
S.~K.~Choi {\it et al.} [Belle Collaboration],
Phys.\ Rev.\ Lett.\  {\bf 91}, 262001 (2003).

\bibitem{Aubert:2005rm}
B.~Aubert \textit{et al.} [BaBar],
Phys. Rev. Lett. \textbf{95} (2005), 142001
[arXiv:hep-ex/0506081 [hep-ex]].


\bibitem{Aubert:2006ge}
B.~Aubert \textit{et al.} [BaBar],
Phys. Rev. Lett. \textbf{98} (2007), 212001
[arXiv:hep-ex/0610057 [hep-ex]].


\bibitem{Belle:2011aa}
  A.~Bondar {\it et al.} [Belle Collaboration],
  Phys.\ Rev.\ Lett.\  {\bf 108}, 122001 (2012).

\bibitem{Aaij:2015tga}
  R.~Aaij {\it et al.} [LHCb Collaboration],
  Phys.\ Rev.\ Lett.\  {\bf 115}, 072001 (2015).

\bibitem{Tornqvist:1993vu}
  N.~A.~Tornqvist,
  On deusons or deuteron - like meson meson bound states,
  Nuovo Cim.\ A {\bf 107}, 2471 (1994).

\bibitem{Tornqvist:1993ng}
  N.~A.~Tornqvist,
  From the deuteron to deusons, an analysis of deuteron-like meson meson bound states,
  Z.\ Phys.\ C {\bf 61}, 525 (1994).

\bibitem{Liu:2008fh}
  Y.~R.~Liu, X.~Liu, W.~Z.~Deng and S.~L.~Zhu,
  Is $X(3872) $ Really a Molecular State?,
  \href{https://link.springer.com/article/10.1140\%2Fepjc\%2Fs10052-008-0640-4}{Eur.\ Phys.\ J.\ C {\bf 56}, 63 (2008)},
  \href{https://arxiv.org/abs/0801.3540}{[arXiv:0801.3540 [hep-ph]]}.

\bibitem{Thomas:2008ja}
  C.~E.~Thomas and F.~E.~Close,
  Is $X(3872)$ a molecule?,
  \href{https://journals.aps.org/prd/abstract/10.1103/PhysRevD.78.034007}{Phys.\ Rev.\ D {\bf 78}, 034007 (2008)},
  \href{https://arxiv.org/abs/0801.3540}{[arXiv:0805.3653 [hep-ph]]}.

\bibitem{Lee:2009hy}
  I.~W.~Lee, A.~Faessler, T.~Gutsche and V.~E.~Lyubovitskij,
  $X(3872)$ as a molecular $D\bar{D}^*$ state in a potential model,
  \href{https://journals.aps.org/prd/abstract/10.1103/PhysRevD.80.094005}{Phys.\ Rev.\ D {\bf 80}, 094005 (2009)},
  \href{https://arxiv.org/abs/0910.1009}{[arXiv:0910.1009 [hep-ph]]}.

\bibitem{Liu:2009ei}
  X.~Liu and S.~L.~Zhu,
  $Y(4143)$ is probably a molecular partner of $Y(3930)$,
  \href{https://journals.aps.org/prd/abstract/10.1103/PhysRevD.80.017502}{Phys.\ Rev.\ D {\bf 80}, 017502 (2009)},
  {Erratum: [Phys.\ Rev.\ D {\bf 85}, 019902 (2012)]},
 \href{https://arxiv.org/abs/0910.1009}{[arXiv:0903.2529 [hep-ph]]}.

\bibitem{Sun:2011uh}
  Z.~F.~Sun, J.~He, X.~Liu, Z.~G.~Luo and S.~L.~Zhu,
  $Z_b(10610)^\pm$ and $Z_b(10650)^\pm$ as the $B^*\bar{B}$ and $B^*\bar{B}^{*}$ molecular states,
  \href{https://journals.aps.org/prd/abstract/10.1103/PhysRevD.84.054002}{Phys.\ Rev.\ D {\bf 84}, 054002 (2011)},
  \href{https://arxiv.org/abs/1106.2968}{[arXiv:1106.2968 [hep-ph]]}.

\bibitem{Chen:2015loa}
  R.~Chen, X.~Liu, X.~Q.~Li and S.~L.~Zhu,
  Identifying exotic hidden-charm pentaquarks,
  \href{https://journals.aps.org/prl/abstract/10.1103/PhysRevLett.115.132002}{Phys.\ Rev.\ Lett.\  {\bf 115}, 132002 (2015)},
  \href{https://arxiv.org/abs/1507.03704}{[arXiv:1507.03704 [hep-ph]]}.

\bibitem{Wigner:1947zz}
E.~P.~Wigner and L.~Eisenbud,
Phys. Rev. \textbf{72} (1947), 29-41

\bibitem{Hale:1987zz}
G.~M.~Hale, R.~E.~Brown and N.~Jarmie,
Phys. Rev. Lett. \textbf{59} (1987), 763-766

\bibitem{Humblet:1991zz}
J.~Humblet, B.~W.~Filippone and S.~E.~Koonin,
Phys. Rev. C \textbf{44} (1991), 2530-2535

\bibitem{Taylor}
J. R. Taylor, Scattering Theory: The Quantum Theory on
Nonrelativistic Collisions (John Wiley $\&$ Sons, New York, 1972).

\bibitem{Hazi}
A. U. Hazi and H. S. Taylor,
Phys. Rev. A \textbf{1}, 1109 (1970).

\bibitem{Kukulin}
V. I. Kukulin, V. M. Krasnoplsky, and J. Horacek,
Theory of Resonances: Principles and Applications (Kluwer, Dordrecht, The Netherlands, 1989).

\bibitem{csm1}
Y. K. Ho,
Phys. Rep. \textbf{99}, 1 (1983).

\bibitem{csm2}
N. Moiseyev,
Phys. Rep. \textbf{302}, 212 (1998).

\bibitem{csm3}
R. A. Weder, J. Math. Phys. \textbf{15}, 20 (1974).

\bibitem{csm4}
A. D. Alhaidari, Phys. Rev. A \textbf{75}, 042707 (2007).

\bibitem{csm5}
M. Bylicki, G. Pestka, and J. Karwowski, Phys. Rev. A \textbf{77},
044501 (2008).

\bibitem{csm6}
J. Y. Guo, X. Z. Fang, P. Jiao, J. Wang, and B. M. Yao, Phys.
Rev. C \textbf{82}, 034318 (2010).

\bibitem{csm7}
Q. Liu, J. Y. Guo, Z. M. Niu, and S. W. Chen, Phys. Rev. C \textbf{86},
054312 (2012).

\bibitem{csm8}
Q. Liu, J. Y. Guo, Z. M. Niu, and S. W. Chen, Phys. Rev. C \textbf{86},
054312 (2012).

\bibitem{csm9}
M. Shi, Q. Liu, Z. M. Niu and  J. Y. Guo,
Phys. Rev. C \textbf{90}, 034319 (2014).


\bibitem{Chen:2017jjn}
R.~Chen, A.~Hosaka and X.~Liu,
Phys. Rev. D \textbf{96} (2017) no.11, 114030
[arXiv:1711.09579 [hep-ph]].



\bibitem{Meguro:2011nr}
  W.~Meguro, Y.~R.~Liu and M.~Oka,
  Possible $\Lambda_c\Lambda_c$ molecular bound state,
  Phys.\ Lett.\ B {\bf 704}, 547 (2011)

\bibitem{Li:2012bt}
  N.~Li and S.~L.~Zhu,
  Hadronic Molecular States Composed of Heavy Flavor Baryons,
  Phys.\ Rev.\ D {\bf 86}, 014020 (2012)

\bibitem{Lee:2011rka}
  N.~Lee, Z.~G.~Luo, X.~L.~Chen and S.~L.~Zhu,
  Possible Deuteron-like Molecular States Composed of Heavy Baryons,
  Phys.\ Rev.\ D {\bf 84}, 014031 (2011)


\bibitem{Riska:1999fn}
  D.~O.~Riska and G.~E.~Brown,
  Two pion exchange interaction between constituent quarks,
  Nucl.\ Phys.\ A {\bf 653}, 251 (1999).

\bibitem{Rijken:1998yy}
  T.~A.~Rijken, V.~G.~J.~Stoks and Y.~Yamamoto,
  Soft core hyperon - nucleon potentials,
  Phys.\ Rev.\ C {\bf 59}, 21 (1999).

\bibitem{Riska:2000gd}
  D.~O.~Riska and G.~E.~Brown,
  Nucleon resonance transition couplings to vector mesons,
  Nucl.\ Phys.\ A {\bf 679}, 577 (2001).

\bibitem{Klempt:2002ap}
  E.~Klempt, F.~Bradamante, A.~Martin and J.~M.~Richard,
  Antinucleon nucleon interaction at low energy: Scattering and protonium,
  Phys.\ Rept.\  {\bf 368}, 119 (2002).

\bibitem{abc}
J. Aguilar and J. M. Combes, Commun. Math. Phys. \textbf{22}, 269
(1971); E. Balslev and J. M. Combes, ibid. \textbf{22}, 280 (1971).

\bibitem{Olive:2016xmw}
  C.~Patrignani {\it et al.} [Particle Data Group],
  Review of Particle Physics,
  Chin.\ Phys.\ C {\bf 40}, 10, 100001 (2016).

\end{thebibliography}
\end{document}